\documentclass[aps,prd,nofootinbib,twocolumn,floatfix,superscriptaddress]{revtex4-1}
\usepackage{graphicx}
\graphicspath{{figures/}} 
\usepackage{times}
\begin{document}
%
%
\title{Coulomb Confinement from the Yang--Mills Vacuum State in 2+1 Dimensions} 
\author{Jeff Greensite}\email{jgreensite@gmail.com}
\affiliation{Physics and Astronomy Dept., San Francisco State
University, San Francisco, CA~94132, USA}
\author{{\v S}tefan Olejn\'{\i}k}\email{stefan.olejnik@savba.sk}
\affiliation{Institute of Physics, Slovak Academy
of Sciences, SK--845 11 Bratislava, Slovakia}
\date{\today}
\begin{abstract}
	The Coulomb-gauge ghost propagator, and the color-Coulomb potential, are computed in an ensemble of configurations derived from our recently proposed Yang--Mills vacuum wavefunctional in $2+1$ dimensions. The results are compared to the corresponding values obtained by standard Monte Carlo simulations in three Euclidean dimensions. The agreement is quite striking for the Coulomb-gauge ghost propagator. The color-Coulomb potential rises linearly at large distances, but its determination suffers from rather large statistical fluctuations, due to configurations with very low values of $\mu_0$, the lowest eigenvalue of the Coulomb-gauge Faddeev--Popov operator. However, if one imposes cuts on the data, effectively leaving out configurations with very low $\mu_0$, the agreement of the potential in both sets of configurations is again satisfactory, although the errorbars grow systematically as the cutoff is eliminated.
\end{abstract}
\maketitle

%
%
\section{Introduction}\label{sec:intro}

	Non-perturbative properties of non-abelian gauge theories, in particular color confinement, chiral symmetry breaking, and the existence of the mass gap, must be encoded in the 
structure of the ground state (vacuum) of these theories.   It is natural, then, to look for evidence of these properties in the vacuum wavefunctional $\Psi_0[A]$ of quantized gauge field theory in some physical gauge.   The simplest non-trivial setting is SU(2) gauge theory with no dynamical matter fields, and in one lower space dimension where the theory is super-renormalizable.

	Recently, we have proposed a very simple approximate form for the Yang--Mills vacuum wavefunctional in temporal gauge and $D=2+1$ dimensions \cite{Greensite:2007ij}:\footnote{Expressions below are assumed to be properly defined on a lattice, with lattice spacing serving as regulator, but for simplicity we will often use continuum notations.}
\begin{eqnarray}
\nonumber
\Psi_0[A]&{=}&\label{eq:ourWF}
\exp\Biggl[-{\textstyle{\frac{1}{2}}}\displaystyle\int d^2xd^2y\\
&\times&B^a(x){\displaystyle
\left(\frac{1}{\sqrt{-{\cal D}^2-
\lambda_0+m^2}}\right)_{xy}^{ab}}B^b(y)\Biggr].
\end{eqnarray}
Here ${\cal D}^2$ is the covariant Laplacian in the adjoint representation, whose lowest eigenvalue is $\lambda_0$,  $m$ is a constant with dimensions of mass proportional to $g^2\sim 1/\beta$,  and ${B^a(x)}=F_{12}^a(x)$ is the color magnetic field strength.   On the lattice, 
$-{\cal D}^2$ is given by 
\begin{eqnarray}
\lefteqn{ \left({-{\cal D}^2}\right)^{ab}_{xy} =} 
\\
& & \displaystyle\sum_{k=1}^2 \left[ 2\delta^{ab}\delta_{xy} -
       {\cal U}^{ab}_k(x)\delta_{y,x+\hat{k}} -{\cal U}^{\dagger ba}_k(x-\hat{k})\delta_{y,x-\hat{k}}\right],
\nonumber
\end{eqnarray}
where the ${\cal U}_k(x)$ are the link fields in the adjoint representation. 

	The wavefunctional in Eq.~(\ref{eq:ourWF}) is reminiscent of Samuel's~\cite{Samuel:1996bt}, the difference being that in his proposal a single free parameter $m^2_0$ replaces our $(-\lambda_0+m^2)$ in the denominator. The reason for subtracting the lowest eigenvalue from the operator   $(-{\cal D}^2)$ is that our numerical simulations indicate that the spectrum of this operator may be divergent in the continuum limit \cite{Greensite:2007ij}.\footnote{It is a little difficult to compare our wavefunctional directly with that of Karabali and Nair \cite{Karabali:1998yq,Karabali:2009rg}, because their proposed wavefunctional in new variables, when converted back to temporal gauge and the usual variables of gauge theory, is not gauge invariant. One must therefore suggest some gauge invariant extension.  The simplest such extension, proposed in \cite{Karabali:1998yq} and investigated numerically in \cite{Greensite:2007ij}, does not seem to give the correct string tension.  Other extensions are, however, possible.}

	The proposed vacuum  wavefunctional (\ref{eq:ourWF}) has quite a few attractive properties:
\begin{enumerate}
\item
In the free-field limit ($g\to 0$), the covariant Laplacian turns into an ordinary Laplacian, $\lambda_0$ and $m$ vanish, and $\Psi_0[A]$ becomes the well-known vacuum wavefunctional of electrodynamics: 
\begin{eqnarray}
\nonumber
\Psi_0[A]&=&
\exp\Bigg\lbrace-{\textstyle{\frac{1}{2}}}\displaystyle\int d^2xd^2y\; 
[\partial_1 A_2^a(x)-\partial_2 A_1^a(x)]\\
&\times&\left(\frac{\delta^{ab}}{\sqrt{-\nabla^2}}\right)_{xy}[\partial_1 A_2^b(y)-\partial_2 A_1^b(y)]\Bigg\rbrace.
\end{eqnarray}
\item 
The expression~(\ref{eq:ourWF}) is a good approximation to the true vacuum also in a completely different corner of the configuration space, namely if we restrict to fields constant in space and varying only in time. In the large-volume limit the solution of the Yang--Mills Schr\"odinger equation in that case is, up to $1/V$ corrections:
\begin{equation}
\Psi_0 =\exp\left[\displaystyle
-{\textstyle{\frac{1}{2}}}gV\frac{(\vec{A_1}\times\vec{A_2})\cdot(\vec{A_1}\times\vec{A_2})}
{\sqrt{\vec{A}_1\cdot\vec{A}_1+\vec{A}_2\cdot\vec{A}_2}}
\right],
\end{equation}
and exactly the same expression follows from (\ref{eq:ourWF}) assuming $\vert g \vec{A}_{1,2}\vert \gg m, \sqrt{\lambda_0}$.
\item 
If we divide the field strength $B^a(x)$ into ``fast'' and ``slow'' components, the part of the (squared) vacuum wavefunctional that depends only on $B_\mathrm{slow}$ is 
\begin{equation}\label{eq:dim_red}
\vert\Psi_0\vert^2\approx\displaystyle\exp\left[
-\frac{1}{m}\int d^2x\; 
B^a_\mathrm{slow}(x)\;B^a_\mathrm{slow}(x)\right].
\end{equation}
Such a form is expected on the basis of dimensional-reduction arguments \cite{Greensite:1979yn,Olesen:1981zp,Ambjorn:1984mb}: it is exactly the probability measure for Yang--Mills theory in two Euclidean dimensions, which (i)~is confining for $m>0$, and (ii) exhibits Casimir scaling for string tensions of all color-charge representations. The fundamental string tension is easily computed as $\sigma_F(\beta)=3 m/(4\beta)$. The last expression can be used to fix the value of $m$ in Eq.~(\ref{eq:ourWF}) at a given $\beta$ from the known value of $\sigma_F(\beta)$. 
\item
Confinement requires $m$ to be larger than zero.  If one takes $m$ in the wavefunctional (\ref{eq:ourWF}) as a variational parameter and computes (approximately) the expectation value of the Yang--Mills Hamiltonian, one finds that a non-zero (and finite) value of $m$ is energetically preferred.
\item 
If we fix the mass $m$ in the wavefunctional to get the right string tension $\sigma_F$ at a given $\beta$, we can test our proposal by calculating \textit{e.g.}\ the mass gap of the theory. We have proposed a recursive procedure for generating independent lattice configurations with the probability distribution given by the square of the wavefunctional (\ref{eq:ourWF}) (see Ref.\ \cite{Greensite:2007ij} and Sec.\ \ref{sec:recursion} for details). We call two-dimensional lattice configurations obtained in this way ``recursion lattices''. One can compute observables with these lattices, and compare the results with corresponding values obtained from ``Monte Carlo lattices'', \textit{i.e.}\ two-dimensional slices of lattices generated in a full $D=3$ lattice Monte Carlo simulation. It turns out that, given the asymptotic string tension as input, the mass gap comes out fairly accurately from our wavefunctional. The discrepancies with the Monte Carlo results of Meyer and Teper \cite{Meyer:2003wx} for the $0^+$ glueball mass are at the level of at most a few ($<6$) percent.
\item 
The dimensional reduction form (\ref{eq:dim_red}) at large distances implies an area law fall-off for large Wilson loops, and also Casimir scaling of higher-representation Wilson loops. The question is then how Casimir scaling turns into $N$-ality dependence at large distances, \textit{i.e.}\ how color screening enters in this setting. There are indications that terms needed for color screening might be contained in our simple wavefunctional and would appear as corrections to the dimensional-reduction form~(\ref{eq:dim_red}).
\end{enumerate}

	In this article we proceed further in comparing quantities derived from our proposed vacuum wavefunctional, with those obtained from full Monte Carlo simulations of the Yang--Mills theory, by computing the Coulomb-gauge ghost propagator and the color-Coulomb potential.  The translation between temporal gauge and the minimal Coulomb gauge was discussed in detail in Ref.\ \cite{Greensite:2004ke}, but the conclusion (\textit{cf.\/}~Sec.~\ref{sec:translation}) is simply that the Yang--Mills wavefunctional in Coulomb gauge can be obtained by restricting the temporal-gauge wavefunctional to transverse gauge fields. In Sec.~\ref{sec:recursion} we review the recursion procedure used to generate lattice configurations with probability weighting given by the square of the wavefunctional~(\ref{eq:ourWF}). Our numerical results are presented in Sec.~\ref{sec:results}, first for the Coulomb-gauge ghost propagator in Sec.~\ref{subsec:ghost}, and then for the color-Coulomb potential in Sec.~\ref{subsec:potential}.  The latter section also contains a discussion of subtleties in the determination of the color-Coulomb potential, due to exceptional lattice configurations. Our conclusions are briefly summarized in Sec.~\ref{sec:conclusions}.  
    
%
%
\section{From temporal to Coulomb gauge}\label{sec:translation}

	The proposed wavefunctional, Eq.~(\ref{eq:ourWF}), is formulated as an approximate solution of the Yang--Mills Schr\"odinger equation in the temporal gauge. To compute quantities in Coulomb gauge we need to find its Coulomb-gauge equivalent. This is an easy task, due to the fact that both the temporal and the Coulomb gauge are compatible with a Hamiltonian formulation and a physical transfer matrix.
	
	In temporal gauge, $A_0=0$, in $D=2+1$ dimensions the continuum Hamiltonian has the simple canonical form
\begin{equation}\label{eq:TGHamiltonian}
H = {\textstyle{\frac{1}{2}}} \int d^2 x \left[ \sum_{i=1}^2 E_i^a(x)^2 + B^{a}(x)^2 \right],
\end{equation}
where the color-electric fields $E_i^a(x)=-i[\delta/\delta A^a_i(x)]$ are operators canonically conjugate to the spatial components of the vector potential. The temporal gauge still possesses invariance under space-dependent, time-independent gauge transformations $\Omega(x)$. The generator of local infinitesimal space-dependent gauge transformations, in the absence of external color sources, is given by
\begin{equation}\label{eq:generator}
G^a(x)=-\sum_{i=1}^2 {\cal{D}}^{ab}_i[A]\;E^b_i(x),
\end{equation}
${\cal{D}}_i[A]$ is the covariant derivative. Physical wavefunctionals are required to satisfy the Gauss law constraint
\begin{equation}\label{eq:Gauss}
G^a(x)\Psi[A]=\sum_{i=1}^2\left( \delta^{ac} \partial_i + g \epsilon^{abc} A^b_i \right) {\delta \over \delta A^c_i }\Psi[A] = 0, 
\end{equation}
which means that the wavefunctional must be invariant under infinitesimal gauge transformations.
 
 Due to the local gauge invariance of wavefunctionals in the temporal gauge, the volume of the gauge group can be extracted
 from inner products of wavefunctionals
\begin{equation}\label{eq:inner}
\langle\Psi_1\vert\Psi_2\rangle=\int[DA]\;\Psi_1^\ast[A]\Psi_2[A]
\end{equation}
via the Faddeev--Popov procedure \cite{Zwanziger:2003cf}. We parametrize configurations $A$ by $A={}^{\Omega^{-1}} A_\perp$, where $A_\perp$ is the representative of $A$ in the minimal Coulomb gauge, \textit{i.e.}\ $A_\perp$ is transverse, $\nabla\cdot A_\perp=0$, and belongs to the fundamental modular region (FMR), $A_\perp\in \Lambda$; $\Omega$~denotes the gauge transformation that brings the configuration $A$ to the minimal Coulomb gauge. The Faddeev--Popov formula then gives
\begin{equation}\label{eq:innerFP}
\langle\Psi_1\vert\Psi_2\rangle=\int_\Lambda\;[DA_\perp]\;\det {\cal{M}}[A_\perp]\;\Psi_1^\ast[A_\perp]\Psi_2[A_\perp],
\end{equation}
where ${\cal{M}}[A_\perp]=-\nabla\cdot {\cal{D}}[A_\perp]$ is the Faddeev--Popov operator, symmetric and positive-definite for $A_\perp\in\Lambda$. The right side of Eq.~(\ref{eq:innerFP}) is the proper expression for the inner product in the minimal Coulomb gauge. This means that in the operator formalism, the minimal Coulomb gauge is a gauge fixing within the temporal gauge of the remnant local gauge invariance. The wavefunctional in Coulomb gauge is the restriction of the wavefunctional in temporal gauge to transverse fields in the FMR:
\begin{equation}\label{eq:WFcoul}
\Psi^{\mathrm{Coulomb}}[A_\perp]=\Psi[A_\perp],\qquad A_\perp\in\Lambda.
\end{equation}
(The conditions $A_0=0$ and $\nabla\cdot A=0$ cannot both be imposed at all times, but can be imposed at one fixed time, and this is all that is required.)

	The vacuum expectation value of an operator $Q$ in Coulomb gauge can then be computed from
\begin{equation}
\langle Q\rangle =\langle\Psi_0^{\mathrm{Coulomb}}\vert Q[A_\perp]\vert\Psi_0^{\mathrm{Coulomb}}\rangle
\end{equation}
Inverting the Faddeev-Popov gauge fixing takes us to
\begin{equation}
\langle Q \rangle = \langle\Psi_0\vert Q\left[{}^\Omega \! A\right]\vert\Psi_0\rangle,
\label{eq:vev}
\end{equation}
\textit{i.e.}\ we generate configurations following the probability distribution $\Psi_0^2$, transform them to the Coulomb gauge, and evaluate the observable $Q$ in the transformed configuration.  From the path-integral representation of the vacuum state, we may also go from (\ref{eq:vev}) to
\begin{equation}
\langle Q \rangle = \left\langle Q\left[{}^{\Omega'}\!\! A({\bf x},t=0)\right] \right\rangle
\end{equation}
where the right hand side is the expectation value obtained in $D=3$ Euclidean dimensions, and $\Omega'$ is the gauge transformation which takes the gauge field on a $t=0$ time-slice into Coulomb gauge. 

%
%
\section{Generation of ``recursion lattices''}\label{sec:recursion}

	To assess the consequences of the proposed wavefunctional, Eq.\ (\ref{eq:ourWF}), one would like to generate lattice configurations with probability distribution given by the square of the wavefunctional\footnote{We have absorbed the coupling $g$ into the definition of $A_k$, which accounts for the factor of $1/g^2$ in the exponent in
Eq.~(\ref{eq:g2}).}
\begin{eqnarray}
\lefteqn{P[A] = \left\vert\Psi_0[A]\right\vert^2} 
\nonumber \\
&=& \exp\left[- {1\over g^2}\int d^2x d^2y ~ B^a(x) {\cal{K}}^{ab}_{xy}[A] B^b(y) \right],
\label{eq:g2}
\end{eqnarray}
where
\begin{equation}\label{eq:kernel}
 {\cal{K}}_{xy}^{ab}[A] = \left({1 \over \sqrt{-{\cal{D}}^2 -  \lambda_0   + m^2}} \right)^{ab}_{xy} 
\end{equation}
We have proposed, in Ref.\ \cite{Greensite:2007ij}, an iterative procedure to achieve this goal. Here we summarize, for completeness and reader's convenience, its most important points.

	We define a probability distribution of gauge fields $A$ with the kernel ${\cal{K}}$ depending on the background field $A'$:
\begin{eqnarray}
&&P\left[A;{\cal{K}}[A']\right] \nonumber\\
&\sim&
\exp\left[-{1\over g^2}\int d^2x d^2y ~ B^a(x) {\cal{K}}^{ab}_{xy}[A'] B^b(y) \right].
\label{eq:PAK}
\end{eqnarray}
Assuming the variance of the kernel ${\cal{K}}$ is small among thermalized configurations gauge-fixed to an appropriate gauge, we can approximate $P[A]$ by
\begin{eqnarray}
P[A]  &\approx& P\left[A,\langle {\cal{K}} \rangle \right]
= P\left[A, \int DA' ~ {\cal{K}}[A'] P[A'] \right]
\nonumber \\
&\approx& \int DA' ~ P\left[A,{\cal{K}}[A']\right] P[A'],
\label{eq:PA}
\end{eqnarray}
and solve the equation iteratively
\begin{eqnarray}
P^{(1)}[A] &=& P\left[A;{\cal{K}}[0]\right],
\nonumber \\
\nonumber\dots &&\\
P^{(n+1)}[A] &=& \int DA' ~ P\left[A;{\cal{K}}[A']\right] P^{(n)}[A'].
\end{eqnarray}    

	In practice, we work on a two-dimensional lattice in an axial gauge ($A_1=0$), which enables one to change variables from $A_1$ and $A_2$ to $B$ cheaply, without introducing a field-dependent jacobian. Initially, we set also $A_2=0$. Then the iterative procedure consists of the following steps:
\begin{itemize}
\item[(i)] Given $A_2$, set $A'_2=A_2$.
\item[(ii)] The probability $P\left[A;K[A']\right]$ is gaussian in $B$, diagonalize $K[A']$ and generate new $B$-field stochastically.
\item[(iii)] From $B$, calculate $A_2$ in axial gauge, and compute everything of interest.
\item[(iv)] Go back to step (i), repeat as many times as necessary.
\end{itemize} 
Lattice configurations generated by this procedure are referred to as ``recursion lattices.'' It turns out that the procedure converges rapidly, after $\cal{O}$(10) iterations (cycles above), and the assumption about a small variance of $\cal{K}$ among configurations is supported \textit{a posteriori} by the absence of large fluctuations of the spectrum of $\cal{K}$ evaluated on individual recursion lattices.   

%
%
\section{Coulomb-gauge results}\label{sec:results}
	Confinement exists, of course, in any gauge, but in some gauges the phenomenon may be easier to understand than in others.
Coulomb gauge has received some attention, following the seminal works of  Gribov \cite{Gribov:1977wm} and Zwanziger \cite{Zwanziger:1998ez}, who argued that the low-lying spectrum of the Faddeev--Popov operator in Coulomb gauge probes properties of non-abelian gauge fields that are crucial for the confinement mechanism. The ghost propagator in Coulomb gauge and the color-Coulomb potential are directly related to the inverse of the Faddeev--Popov operator, and play a role in various confinement scenarios. In particular, the color-Coulomb potential represents an upper bound on the physical potential between a static quark and antiquark, which means that a confining color-Coulomb potential is a necessary condition to have a confining static quark potential \cite{Zwanziger:2002sh}. It is therefore an important check on the validity of our proposed vacuum wavefunctional, to see how well it can reproduce the values of Coulomb-gauge observables, such as the ghost propagator and color-Coulomb potential, that can be obtained by standard lattice Monte Carlo techniques.  
	
\subsection{Ghost propagator}\label{subsec:ghost}
	The ghost propagator in Coulomb gauge is given by the inverse of the Faddeev--Popov operator:
\begin{eqnarray}
\nonumber
G(R)&=&\left.\left\langle\left({\cal{M}}[A]^{-1}\right)^{aa}_{xy}\right\rangle\right\vert_{\vert x-y\vert=R}\\
&=&\left.\left\langle\left(-\frac{1}{\nabla\cdot{\cal{D}}[A]}\right)^{aa}_{xy}\right\rangle\right\vert_{\vert x-y\vert=R}.
\label{eq:ghost_prop}
\end{eqnarray}
In $D=2+1$ dimensions, with SU(2) gauge group and lattice link matrices parametrized by
\begin{equation}
U_\mu(x)=b_\mu(x)\mathbf{1}+i \sigma^c a^c_\mu(x),\quad b_\mu(x)^2+a^c_\mu(x)^2=0,
\end{equation}
the lattice Faddeev--Popov operator is given by (assuming the lattice version of the Coulomb-gauge condition $\nabla\cdot A=0$ is satisfied)
\begin{eqnarray}
\nonumber
  {\cal{M}}^{ab}_{xy}&=& \delta^{ab} \sum_{k=1}^2\left\{ \delta_{xy}  \left[b_k(x)
    + b_k(x-\hat{k})\right]\right.\\
    &-& \left. \delta_{x,y-\hat{k}} b_k(x)
    - \delta_{y,x-\hat{k}} b_k(y) \right\}\label{eq:FP}\\
\nonumber
    &-& \epsilon^{abc} \sum_{k=1}^3\left\{ \delta_{x,y-\hat{k}} a^c_k(x)
                  - \delta_{y,x-\hat{k}} a^c_k(y)  \right\}. 
\end{eqnarray}
This operator is symmetric and positive-definite anywhere in the Coulomb-gauge Gribov region, \textit{i.e.}\ for configurations that are local minima of the quantity
\begin{equation}
{\cal{R}}[U]=-\sum_x\sum_{k=1}^2 \mbox{Tr}[U_k(x)]. 
\label{eq:Rcoul}
\end{equation}
More precisely, on a lattice with periodic boundary conditions the Faddeev--Popov operator possesses three trivial zero eigenvalues, the eigenmodes being independent of $x$. The existence of these modes is a consequence of the fact that, even apart from Gribov copies, the Coulomb-gauge condition does not fix the gauge completely.  There is a remnant global symmetry such that if a set of link matrices $U_k(x)$ satisfies the condition, so does $U_k'(x)=\Omega\;U_k(x)\;\Omega^\dagger$, where $\Omega$ is a space-independent SU(2) group element. The inversion of the Faddeev--Popov operator, Eq.~(\ref{eq:FP}), needed to compute the ghost propagator (\ref{eq:ghost_prop}), is therefore performed in the subspace orthogonal to the trivial zero eigenmodes.

	At lattice coupling $\beta=6$ on a $24^2$ lattice, and $\beta=9$ on a $32^2$ lattice,  we have computed the Coulomb-gauge ghost propagator separately in two ensembles of lattice configurations generated in two different ways:
\begin{itemize}
\item[(i)] \textit{recursion lattices} are generated by the procedure described in Section \ref{sec:recursion}; the mass parameter $m$ in the wavefunctional (\ref{eq:ourWF}), at each lattice coupling, was chosen to reproduce the value of the string tension at the same coupling as given by Ref.\ \cite{Meyer:2003wx};
\item[(ii)] \textit{Monte Carlo lattices} are generated by Monte Carlo simulations of euclidean SU(2) lattice gauge theory in $D=3$ dimensions with the standard Wilson action.  From each thermalized configuration only a single (random) space slice at fixed euclidean time was chosen.  The probability measure for such lattice time-slices, when transformed to Coulomb gauge, is given by the true Coulomb-gauge vacuum wavefunctional (i.e.\ the
vacuum state of the corresponding lattice transfer matrix).
\end{itemize}
There were 1000 lattice configurations in each ensemble, at each of the couplings $\beta=6$ and $9$. 
	
	Each two-dimensional lattice configuration was fixed to the Coulomb gauge by minimizing the quantity $\cal{R}$, Eq.~(\ref{eq:Rcoul}), via the usual (over)relaxation method.\footnote{It should be noted that this procedure returns a gauge copy in the Gribov region; no procedure for fixing to the fundamental modular region exists.  However, we feel that there is nothing sacred (or more physical) about Gribov copies in the fundamental modular region, and will be satisfied with local, rather than global minima of $\cal{R}$.} Then, the inverse of the Faddeev--Popov operator was computed using the standard linear-algebra tools (\texttt{Octave}). This enabled us to determine the ghost propagator directly in the coordinate representation, in contrast to most other lattice investigations which generally determine propagators in the momentum representation (see \textit{e.g.\/} Refs.\ \cite{Quandt:2008zj,Nakagawa:2009zf}).

\begin{figure*}[bth!]
\centering
\begin{tabular}{c c}
\includegraphics[width=0.48\hsize]{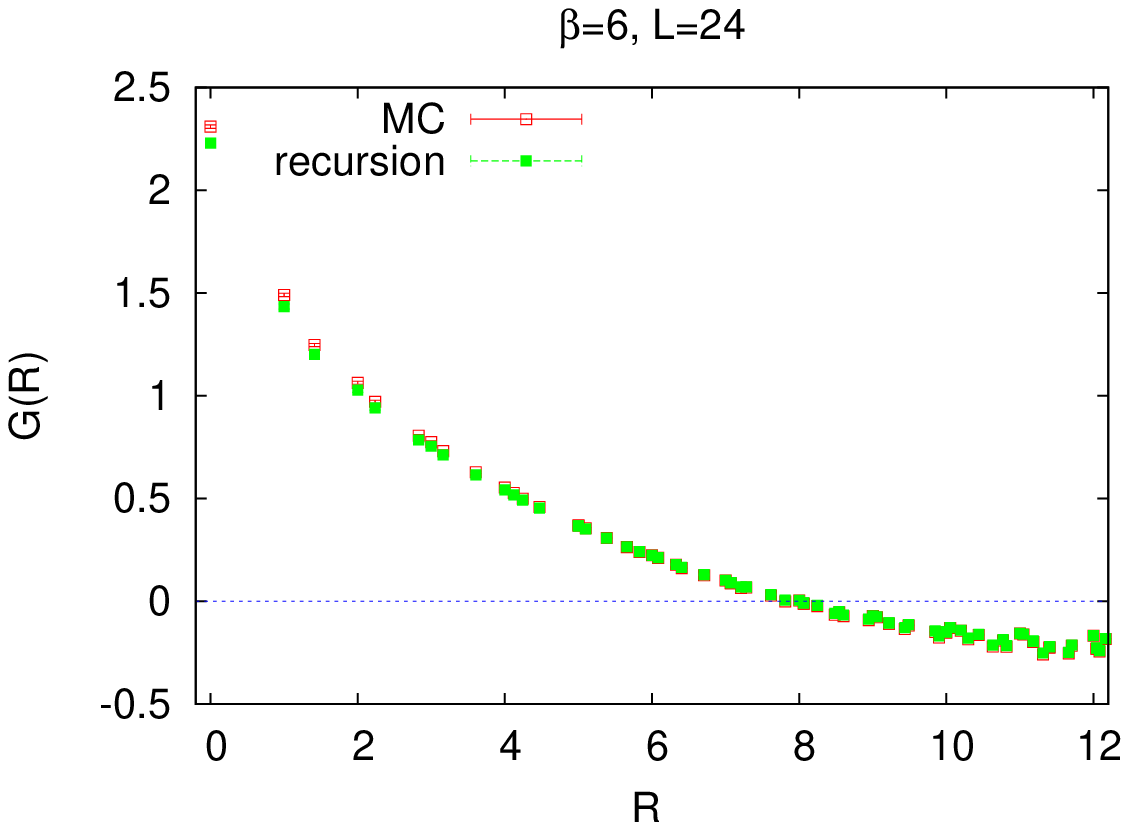} &
\includegraphics[width=0.48\hsize]{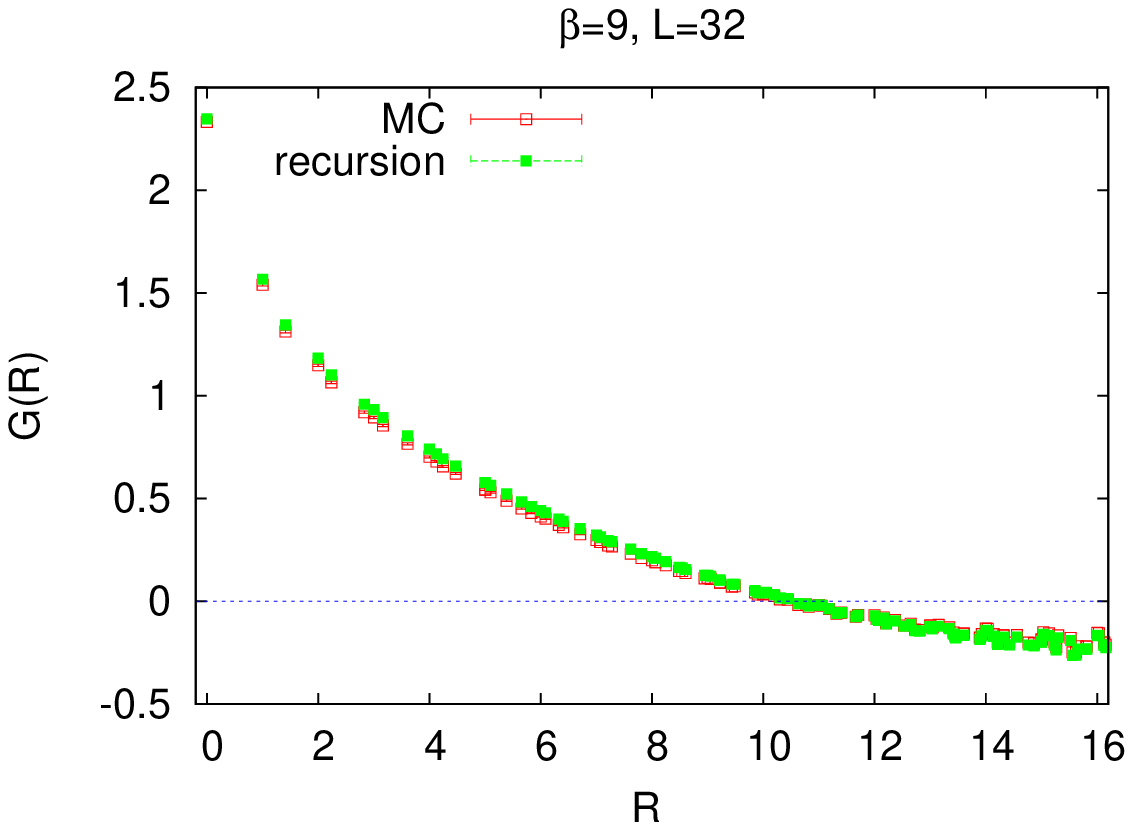}
\end{tabular}
\caption{The Coulomb-gauge ghost propagator: (left) $\beta=6$ on $24^2$ lattice, (right) $\beta=9$ on $32^2$ lattice.}
\label{fig:propagator}
\end{figure*}

	The results for our largest coupling ($\beta=9$) and lattice size ($32^2$) are displayed in Fig.~\ref{fig:propagator} (right). The agreement between recursion and Monte Carlo lattices is really quite striking, with the differences between the displayed data sets being comparable to the size of the symbols. The same agreement is observed also for $\beta=6$ on $24^2$ lattice, \textit{cf.\/} Fig.~\ref{fig:propagator} (left).

\subsection{Color-Coulomb potential}\label{subsec:potential}
\begin{figure*}[tbh!]
\centering
\begin{tabular}{c c}
\includegraphics[width=0.48\hsize]{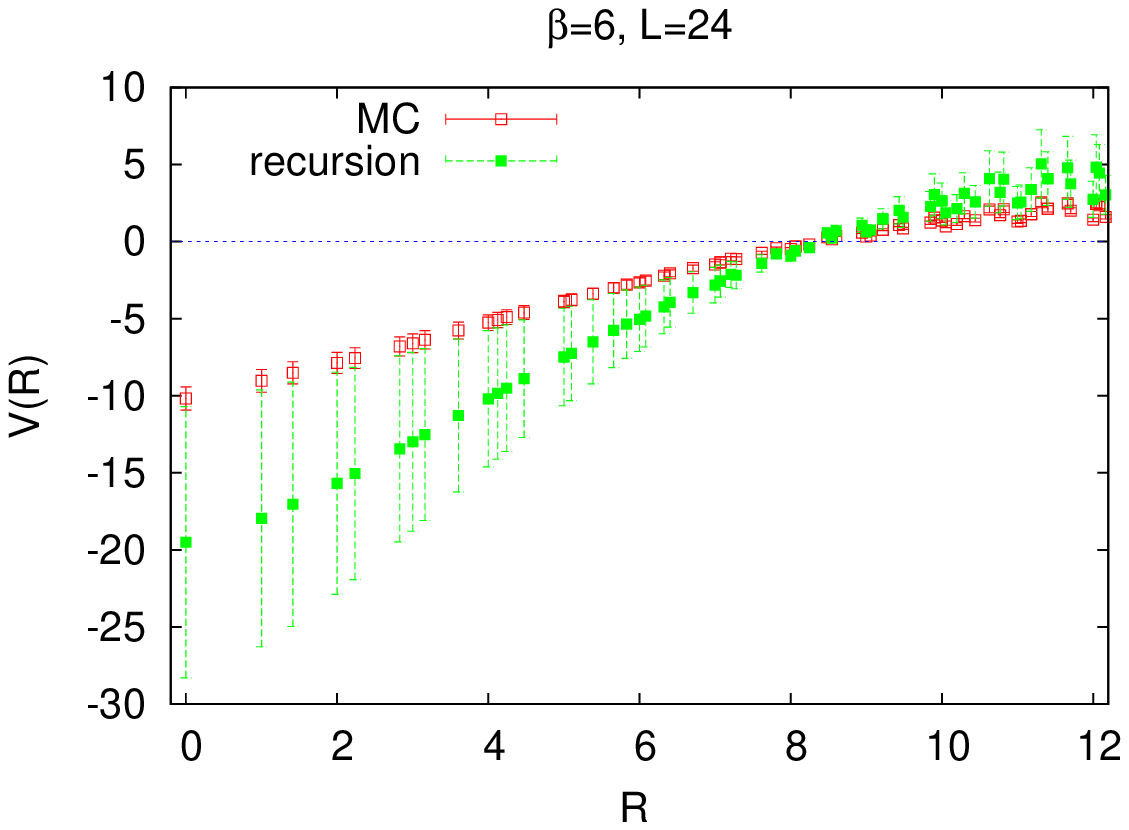} &
\includegraphics[width=0.48\hsize]{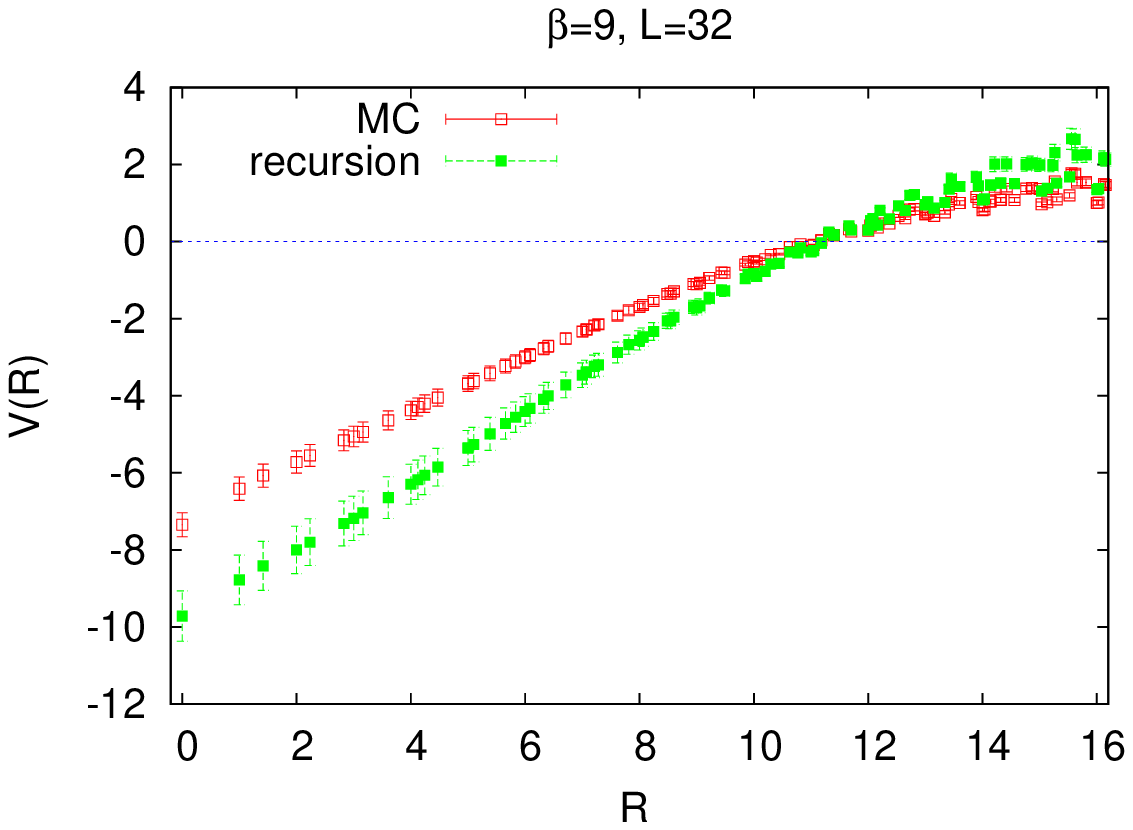}
\end{tabular}
\caption{The color-Coulomb potential computed from all measured configurations: (left) $\beta=6$ on $24^2$ lattice, (right) $\beta=9$ on $32^2$ lattice.}
\label{fig:potential}
\end{figure*}

	The potential between a static quark and antiquark located at points $x$ and $y$, respectively, is proportional to:\footnote{We have omitted below the normalization factor $g^2 C_F/d_A$, where $C_F$ is the eigenvalue of the quadratic Casimir operator in the fundamental color representation and $d_A$ is the dimension of the adjoint representation of the color group, and nevertheless call the above quantity the color-Coulomb potential. The normalization factor would be needed if we intended to get a numerically accurate value of the Coulomb string tension, which was not our goal in the present study.}
\begin{eqnarray}
V(R)&=&\left.-\left\langle\left({\cal{M}}[A]^{-1}(-\nabla^2){\cal{M}}[A]^{-1}\right)^{aa}_{xy}\right\rangle\right\vert_{\vert x-y\vert=R}\\
\nonumber
&=&\left.-\left\langle\left(\frac{1}{\nabla\cdot{\cal{D}}[A]}(-\nabla^2)\frac{1}{\nabla\cdot{\cal{D}}[A]}\right)^{aa}_{xy}\right\rangle\right\vert_{\vert x-y\vert=R}.
\label{eq:Coulomb_potential}
\end{eqnarray}
Provided we know the inverse of the Faddeev--Popov operator in a configuration (on the subspace orthogonal to trivial zero modes, see the preceding section), the computation of the potential is quite straightforward. The result is shown in Fig.~\ref{fig:potential}.

	After what we have seen in Section \ref{subsec:ghost}, the figures for potentials come as a surprise. In the case of the ghost propagator, there was almost no difference between recursion and Monte Carlo lattice ensembles. Now we observe quite strong differences. The origin of these differences can fortunately be identified. Both in the Monte Carlo and recursion ensemble there exist ``exceptional'' configurations in which the lowest nontrivial eigenvalue of the Faddeev--Popov operator $\mu_0$ is still positive, but extremely small. It means that in the space of gauge-equivalent configurations there exist a valley along which the minimized quantity $\cal{R}$, Eq.~(\ref{eq:Rcoul}), almost does not change. Consequently, these configurations were always rather difficult to gauge-fix to Coulomb gauge.\footnote{A similar type of ``exceptional'' hard-to-gauge-fix configurations was encountered \textit{e.g.\/} by the Berlin group\ \cite{Voigt:2008rr}.} The existence of these exceptional configurations does not have a crucial impact on the ghost propagator, since its definition contains a single power of the inverse Faddeev--Popov operator, but their influence is strongly amplified in color-Coulomb potentials, where the inverse appears twice.

	We will illustrate this point on the data for $\beta=9$ on $32^2$ lattice. One can evaluate the values of the potential in a single lattice configuration. The lowest eigenvalue of the Faddeev--Popov operator directly influences the absolute value of the potential at the origin, $\vert V(0)\vert$, so one can classify configurations by their values of $\vert V(0)\vert$, and evaluate average potentials from sets of configurations satisfying a number of cuts: $\{\vert V(0)\vert<\kappa_i, i=1,2,\dots, K\}$.

\begin{figure*}[bth!]
\centering
\begin{tabular}{c c}
\includegraphics[width=0.48\hsize]{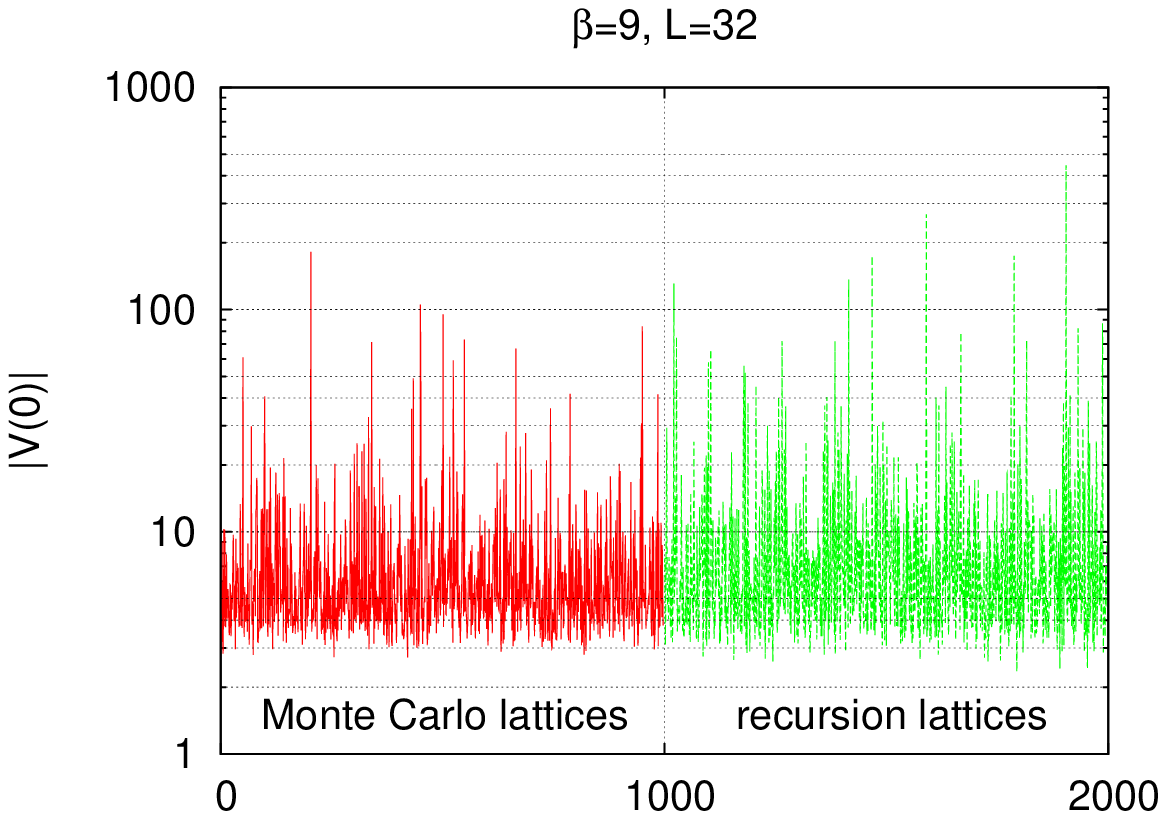}&
\includegraphics[width=0.48\hsize]{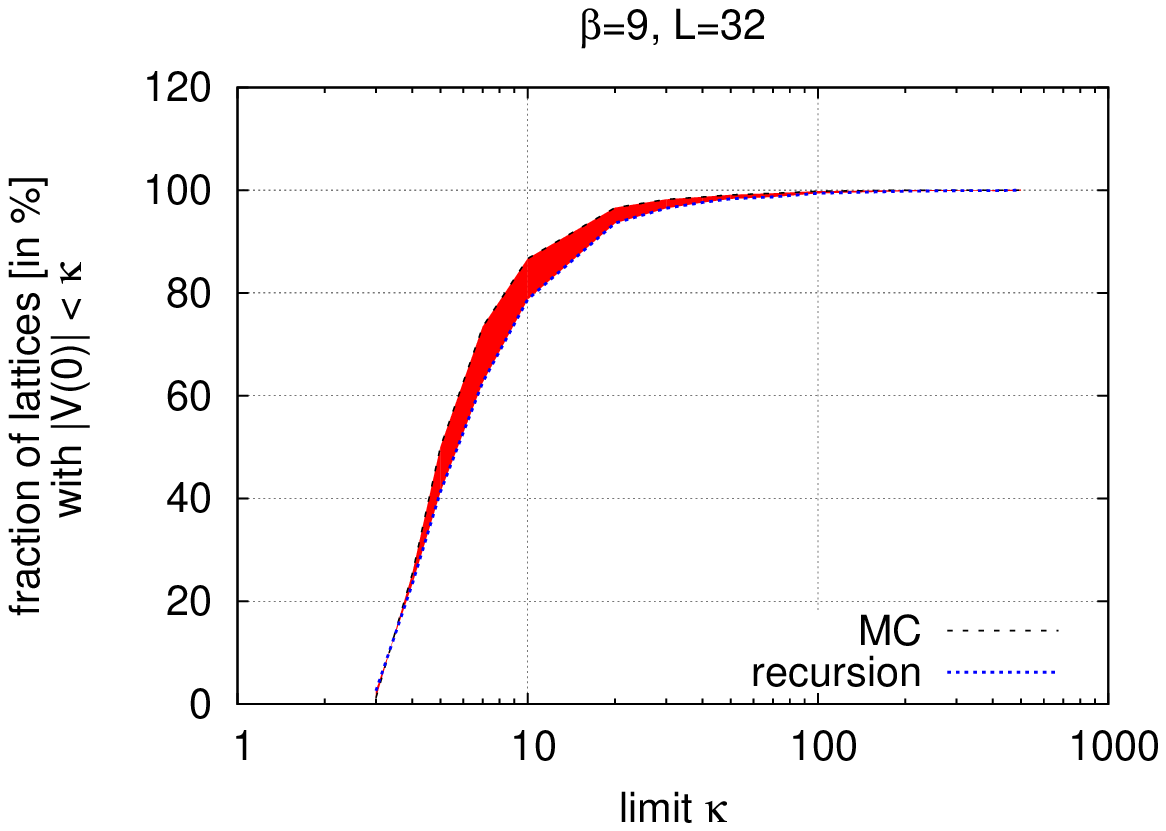}
\end{tabular}
\caption{Left: $\vert V(0)\vert$ in the individual configurations from the recursion and Monte Carlo ensembles. Right: The fraction of configurations with $\vert V(0)\vert<\kappa$. The top curve is for Monte Carlo lattices, the bottom one for recursion lattices. The area between the curves is colored. }
\label{fig:v0_b9_l32b}
\end{figure*}

	Figure \ref{fig:v0_b9_l32b}	(left) displays values of $\vert V(0)\vert$ in the individual configurations from the recursion and Monte Carlo ensembles. One can see that the majority of configurations lie in the range between about 2 to 20, but there are rare instances of configurations above 100 (among Monte Carlo lattices) or even 400 (among recursion lattices). The frequency of these ``exceptional'' configurations in the ensemble is very small, but such configurations give rise to tremendous fluctuations in the measured average values of the potential. The distribution of configurations according to $\vert V(0)\vert$ is further illustrated in Fig.~\ref{fig:v0_b9_l32b} (right).

\begin{figure*}[bth!]
\centering
\begin{tabular}{c c}
\includegraphics[width=0.48\hsize]{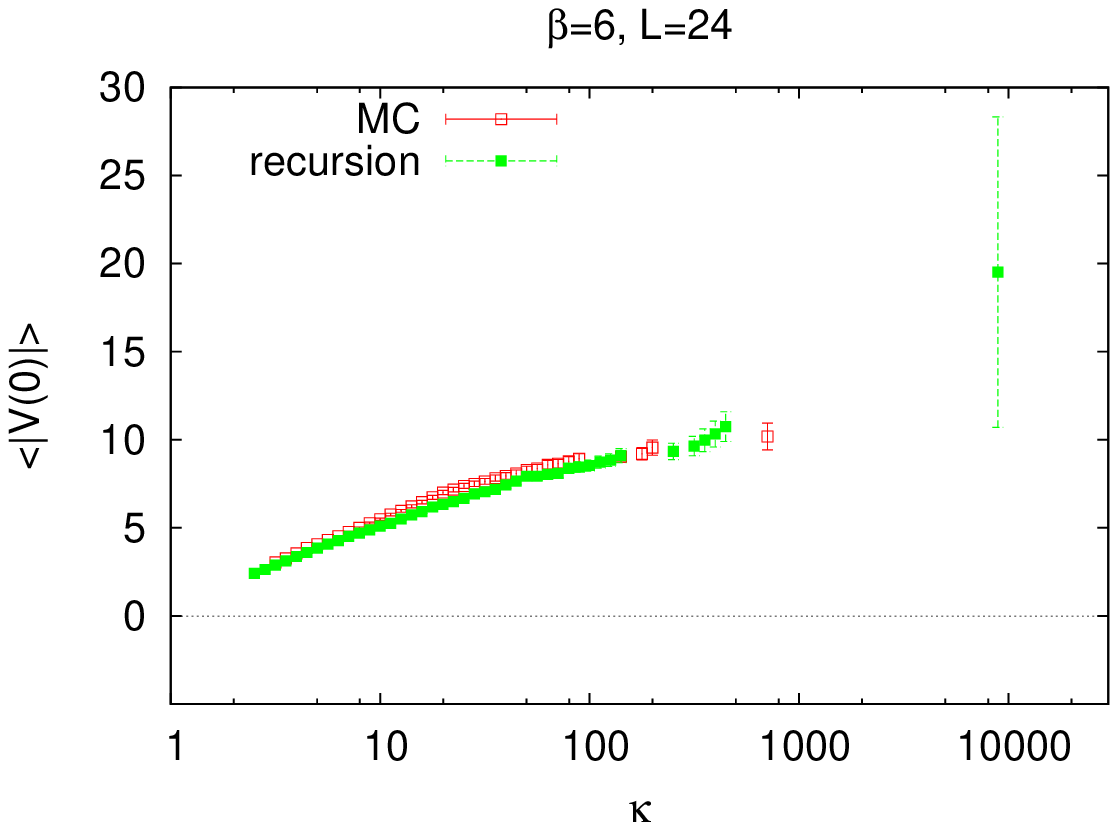}&
\includegraphics[width=0.48\hsize]{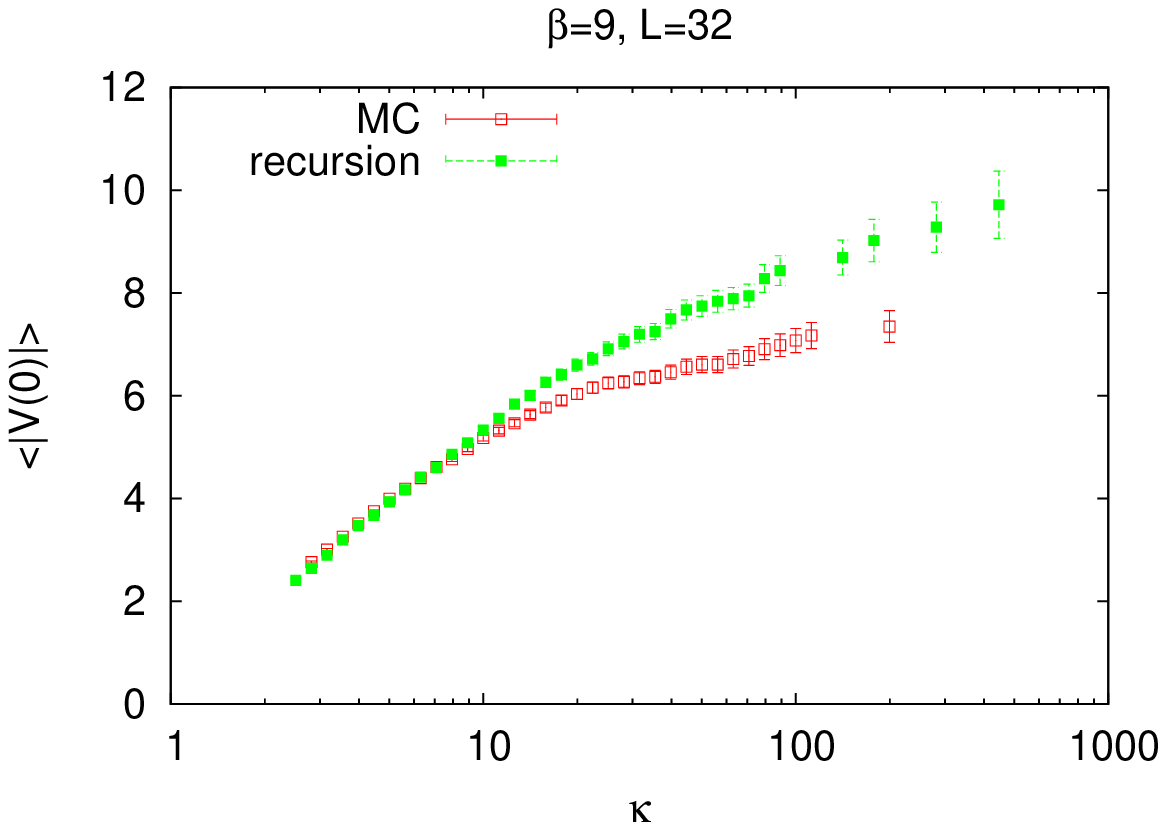}
\end{tabular}
\caption{The average of $\vert V(0)\vert$ evaluated in subsets of configurations satisfying the condition $\vert V(0)\vert<\kappa$: (left) $\beta=6$ on $24^2$ lattice, (right) $\beta=9$ on $32^2$ lattice.}
\label{fig:v0_vs_kappa}
\end{figure*}

	The $\kappa$-dependence of the average value of the magnitude of the color-Coulomb potential $V(R)$ at $R=0$, evaluated in subsets of Monte Carlo and recursion lattices with $\vert V(0)\vert<\kappa$, is displayed in Fig.~\ref{fig:v0_vs_kappa}. At $\beta=9$ (right panel), the average values in both ensembles agree reasonably at least for $\kappa\le{\cal{O}}(10)$. Such a cut on $\vert V(0)\vert$ is satisfied by about 85\% Monte Carlo lattices, and almost 80\% recursion lattices (\textit{cf.\/} Fig.~\ref{fig:v0_b9_l32b}, right). The severity of the problem of ``exceptional'' configurations is seen even more dramatically in the data for $\beta=6$ on $24^2$ lattice (Fig.~\ref{fig:v0_vs_kappa}, left panel). Here the close agreement of the average $\vert V(0)\vert$ between Monte Carlo and recursion lattices persists for the cut-off $\kappa$ at least ${\cal{O}}(100)$ (satisfied by 99\% of lattices); then, a single recursion lattice with extremely high value of $\vert V(0)\vert$ (\textit{i.e.}\ extremely small lowest eigenvalue of the Coulomb-gauge Faddeev--Popov operator) completely distorts the picture, and causes the disagreement between color-Coulomb potentials and the huge errorbars seen in Fig.~\ref{fig:potential}.
	
\begin{figure*}[bth!]
\centering
\begin{tabular}{c c}
\includegraphics[width=0.48\hsize]{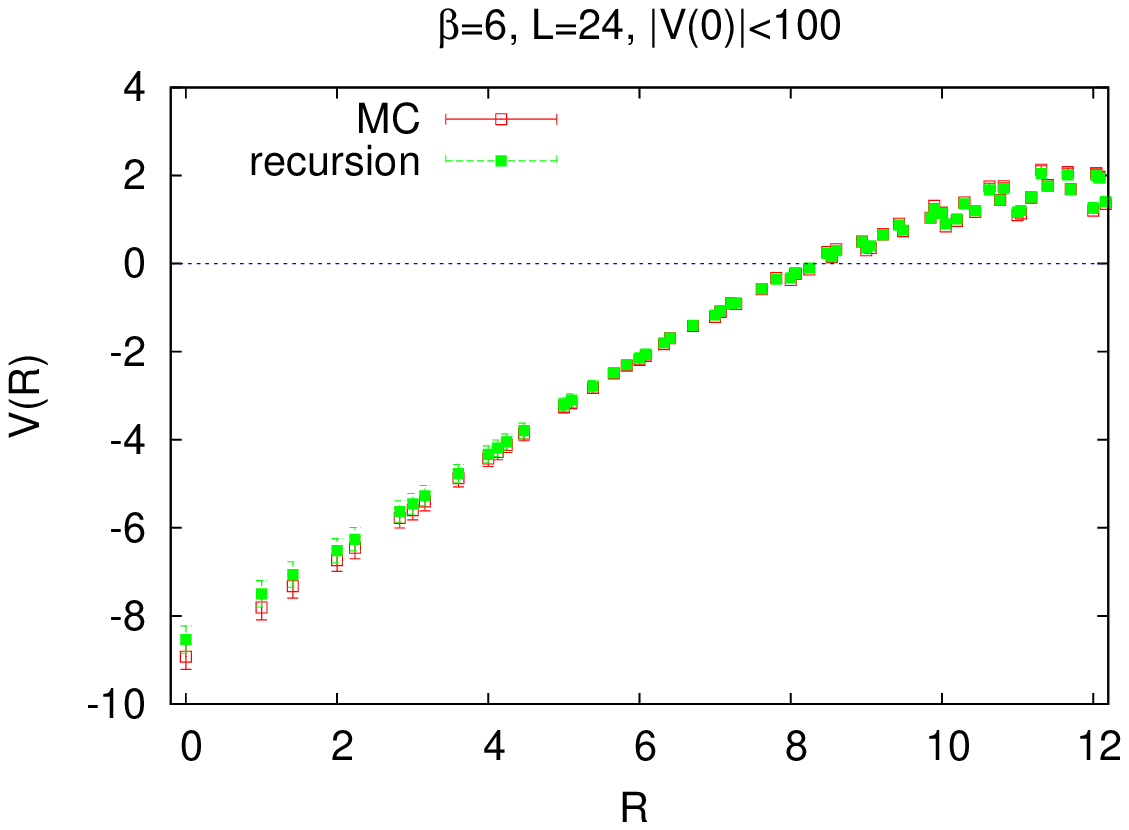}&
\includegraphics[width=0.48\hsize]{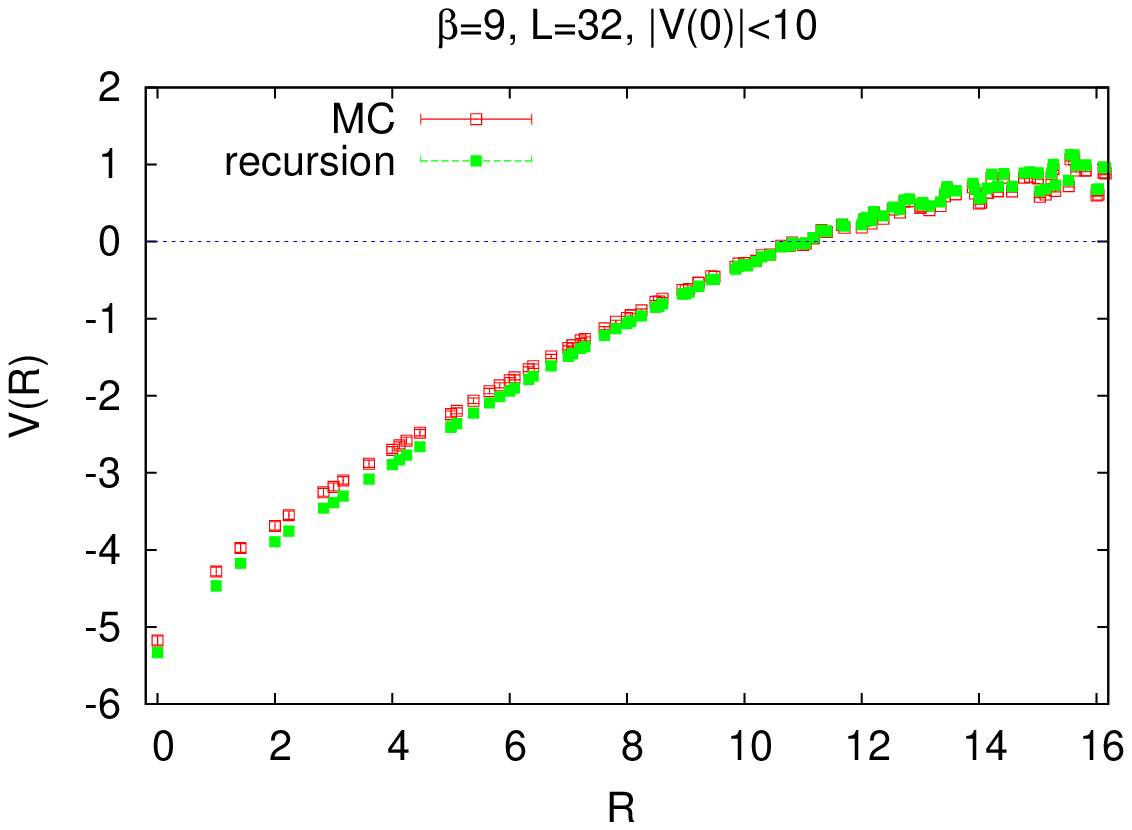}
\end{tabular}
\caption{The color-Coulomb potential computed from configurations with cuts: (left) $\beta=6$ on $24^2$ lattice, $\kappa=100$, (right) $\beta=9$ on $32^2$ lattice, $\kappa=10$.}
\label{fig:potential_b9_l32_cuts}
\end{figure*}

	If the color-Coulomb potential is evaluated in Monte Carlo and recursion lattices with the same (not too high) cuts applied in both ensembles, the fluctuations due to rare configurations, observed in Fig.~\ref{fig:potential}, are tamed. In Fig.~\ref{fig:potential_b9_l32_cuts} we show the results for $\beta=6$ with $\kappa=100$, and for $\beta=9$ with $\kappa=10$. The potentials are linearly rising over a certain range of distances, and agree quite well in recursion and Monte Carlo lattices. The agreement deteriorates somewhat when the value of the cut $\kappa$ is increased, but we believe that approximate agreement would be restored with sufficient (but obviously huge) statistics if the sets of configurations in high $\vert V(0)\vert$ bins were equally populated.
%

%
%
\section{Conclusions}\label{sec:conclusions}    

	The results presented in this paper strengthen our confidence that the vacuum wavefunctional for the temporal-gauge SU(2) Yang--Mills theory in $D=2+1$ dimensions, Eq.~(\ref{eq:ourWF}), is a fairly good approximation to the true ground state of the theory. The evidence supplied by Ref.\ \cite{Greensite:2007ij} and summarized in Section\ \ref{sec:intro} has been now augmented by measurement of two quantities which play a crucial role in understanding confinement in Coulomb gauge:
\begin{enumerate}
\item
	The ghost propagator in Coulomb gauge, computed from an ensemble of recursion lattices (derived from our approximate wavefunctional), and from an ensemble of Monte Carlo lattices, comes out to be practically identical in both ensembles.
\item
	If the same cuts on ``exceptional'' configurations are applied in both ensembles, then the color-Coulomb potential from recursion lattices is also very close to that determined from Monte Carlo lattices. Both potentials are linearly rising over a range of distances (until the effects of lattice periodicity become important). However, one would need to considerably increase the statistics to ensure approximately equal population of exceptional configurations in both ensembles, to convincingly prove that the deviation between potentials in these two ensembles stays tiny even after the cuts are removed.
\end{enumerate}
While we do not claim that  Eq.~(\ref{eq:ourWF}) is exact (it is surely only an approximation to the true ground state), it has now passed a number of very non-trivial tests.  More tests are in progress, and will be reported on at a later time.

     The extension of our proposed vacuum wavefunctional to $D=3+1$ dimensions is straightforward; it simply involves replacing the product
$B^a(x) B^b(y)$ in Eq.~(\ref{eq:ourWF}) by $F_{ij}^a(x) F_{ij}^b(y)$, and replacing two-dimensional integrations by three-dimensional integrals.  It remains true in $3+1$ dimensions that the resulting wavefunctional is exact in the free-field limit, and approximately solves the Yang--Mills Schr\"odinger equation in the zero-mode, strong-field limit (see Ref.\ \cite{Greensite:2007ij} for details). However, going to three space dimensions brings up complications associated with the Bianchi constraint.  Because of that constraint,  numerical simulation of the wavefunctional along the lines discussed in Section \ref{sec:recursion} is much more challenging in $D=3+1$ than in $D=2+1$, and a different approach to testing our proposal is probably required.\footnote{For some recent work using the dimensionally-reduced form of the 3+1 dimensional vacuum wavefunctional, \textit{cf.\/}
Ref.~\cite{Quandt:2010yq}.}

%
%
\acknowledgments{This research is supported in part by the U.S.\ Department of Energy under Grant No.\ DE-FG03-92ER40711 (J.G.), the Slovak Grant Agency for Science, Project VEGA No.\ 2/0070/09, by ERDF OP R\&D, Project CE QUTE ITMS 26240120009, and via CE SAS QUTE  (\v{S}.O.).}       
 
%
%

%

\end{document}